\documentclass[aps,prl,twocolumn,superscriptaddress]{revtex4-1}
\usepackage{graphicx}
\usepackage{epstopdf}
\usepackage{bm}
\usepackage{amssymb}
\usepackage{amsmath}
\usepackage[colorlinks=true]{hyperref}

\begin{document}
\title{Gauge fields and related forces in antiferromagnetic solitons}
\author{Sayak Dasgupta}
\affiliation{Department of Physics and Astronomy, 
	Johns Hopkins University, 
	Baltimore, Maryland 21218, USA
	}
\author{Se Kwon Kim}
\affiliation{Department of Physics and Astronomy, 
	University of California, 
	Los Angeles, California 90095, USA
	}
\author{Oleg Tchernyshyov}
\affiliation{Department of Physics and Astronomy, 
	Johns Hopkins University, 
	Baltimore, Maryland 21218, USA
	}

\begin{abstract}
We derive equations of motion for topological solitons in antiferromagnets under the combined action of perturbations such as an external magnetic field and torque-generating electrical current. Aside from conservative forces, such perturbations generate an effective ``magnetic field'' exerting a gyrotropic force on the soliton and an induced ``electric field'' if the perturbation is time-dependent. We apply the general formalism to the cases of a domain wall and of a vortex. An antiferromagnetic vortex can be effectively moved by combined applications of a magnetic field and an electric current. 
\end{abstract}

\maketitle

Topological solitons and their dynamics have long attracted the attention of physicists because of both fundamental interest and technological applications \cite{PhysRep.194.117}. For example, domain walls to a large extent define the properties of permanent magnets as they mediate the process of magnetization reversal. Recent proposals of using domain walls \cite{Science.320.5873} and skyrmions \cite{NatNano.8.899} for storing and processing digital information have generated a large body of theoretical and experimental works elucidating the properties of these solitons. Their dynamics in ferromagnets is dominated by gyroscopic effects because they are made of little gyroscopes---electron spins. Thus, to propel a ferromagnetic vortex in the $x$ direction of the $xy$ plane, one needs to apply a force in the $y$ direction. Similarly, applying a force to a domain wall in a uniaxial ferromagnet primarily generates its precession. To propel it forward, one has to apply a torque to it. 

A promising new direction of basic and applied research in spintronics is the study of solitons in antiferromagnets. Potential advantages of antiferromagnets are the absence of long-range stray magnetic fields and associated harmful crosstalk, the suppression of gyroscopic effects, and generally faster dynamics \cite{NatNano.11.231}. At the same time, there are new challenges. How does one apply a force to an antiferromagnetic soliton? An external magnetic field couples to the net magnetic moment, which is strongly suppressed in an antiferromagnet. Spin torque couples to the wrong channel, generating rotational, rather than translational, motion of an antiferromagnetic domain wall. 

To generate a net force on an antiferromagnetic soliton, one may follow a general strategy of combining two or more external perturbations. For example, the application of an electric field breaks the equivalence of the two magnetic sublattices in the magnetoelectric antiferromagnet Cr$_2$O$_3$ \cite{JPhysChemSol.4.241}. The order parameter---staggered magnetization---is accompanied by small uniform magnetization parallel to it. The system is a weak ferromagnet, so that an applied magnetic field exerts a force on a domain wall and propels it forward \cite{ApplPhysLett.108.132403}. 

The main goal of this paper is to present a framework for computing the net force on an antiferromagnetic soliton under the action of combined perturbations. We use the Lagrangian formalism for two magnetization fields appropriate for a two-sublattice antiferromagnet, the dominant field of staggered magnetization $\mathbf n(\mathbf r)$ and the subleading field of uniform magnetization $\mathbf m(\mathbf r)$. The advantage of the Lagrangian approach over the standard treatment at the level of the Landau-Lifshitz equations is the ease of calculating the net force acting on a soliton, either through the energy-momentum tensor, or by restricting the Lagrangian of the field $\mathbf n(\mathbf r)$ to a set of collective coordinates such as the soliton's center of mass. Aside from well understood inertial, conservative, and dissipative forces \cite{PhysRevLett.110.127208, PhysRevB.90.104406}, a moving soliton experiences a Lorentz force from an emergent electromagnetic field, an analog of the gyrotropic force in a ferromagnet, and a Magnus force from a passing electric current. 

We discuss two specific examples. For a domain wall in an easy-axis antiferromagnet, we derive the dynamics under the influence of an external magnetic field and of adiabatic spin torque. It is well known \cite{PhysRevLett.106.107206, ApplPhysLett.109.122404} that these perturbations alone are not able to propel a domain wall in an antiferromagnet. We use this familiar example to illustrate the general formalism and to make a few relevant generalizations. We then consider the case of a vortex in an easy-plane antiferromagnet, which has not received as much attention. We show that, in the presence of an external magnetic field $\mathbf H$ along the hard axis, an electric current generates a Magnus force through adiabatic spin torque. For a vortex line in three dimensions, the Magnus force is proportional to the drift velocity of spin current $\mathbf u$: 
\begin{equation}
\mathbf F = 2\pi n \rho \gamma H \int \mathbf u \times d \mathbf r ,
\label{eq:Magnus-force}
\end{equation}
which is analogous to the Magnus force on a vortex line in superconductors in the presence of an electric current \cite{Rev.Mod.Phys.36.45}. Here $\rho$ is the density of inertia of staggered magnetization and $\gamma$ is the gyromagnetic ratio. It is remarkable that the Magnus force (\ref{eq:Magnus-force}) depends on the vortex winding number $n$ but not on its detailed structure. We trace its origin to a subtle change in the topology of the antiferromagnetic vortex. A topological nature of the Magnus force guarantees its robustness. The magnitude and the sign of the Magnus force can be controlled by tuning the applied magnetic field. 

\emph{General formalism.} 
A continuum theory of a collinear antiferromagnet with two sublattices operates with two slowly varying fields $\mathcal M \mathbf m_1(\mathbf r)$ and $\mathcal M \mathbf m_2(\mathbf r)$, where $\mathcal M$ is the magnetization length and $\mathbf m_1$ $\mathbf m_2$ are unit vector fields. In a state of equilibrium, $\mathbf m_1(\mathbf r) = - \mathbf m_2(\mathbf r)$. More generally, the two sublattice fields are expressed in terms of dominant staggered magnetization $\mathbf n = (\mathbf m_1 - \mathbf m_2)/2$ and small uniform magnetization $\mathbf m = \mathbf m_1 + \mathbf m_2$. The constraints $|\mathbf m_1|^2 = 1$ and $|\mathbf m_2|^2 = 1$ translate into  
\begin{equation}
\mathbf m \cdot \mathbf n = 0, 
\quad
|\mathbf n|^2 = 1 - |\mathbf m|^2/4 \approx 1;
\label{eq:constraints}
\end{equation}
the last approximation is valid as long as $|\mathbf m|^2 \ll 1$. 

The dynamics of magnetization fields $\mathbf n$ and $\mathbf m$ is determined by the Lagrangian density 
\begin{equation}
\mathcal L(\mathbf n, \mathbf m) = 
	\mathcal J \mathbf m \cdot (\dot{\mathbf n}\times \mathbf n)
	+ \mathcal M \mathbf m \cdot \mathbf h 
 	- \frac{|\mathcal M\mathbf m|^2}{2\chi}
	- \mathcal U(\mathbf n).
\label{eq:L-n-m}
\end{equation}
Eq.~(\ref{eq:L-n-m}) can be thought of as a Taylor series in powers of $\mathbf m$. The kinetic term $\mathcal J \mathbf m \cdot (\dot{\mathbf n}\times \mathbf n)$ represents the spin Berry phase and plays a crucial role in shaping up the dynamics of magnetization \cite{PhysRevLett.74.1859, PhysRevB.90.104406, PhysRevB.93.104408}; $\mathcal J$ is the density of angular momentum for one sublattice. The term $\mathcal M \mathbf m \cdot \mathbf h$ represents potential energy terms linear in $\mathbf m$, (e.g., the Zeeman coupling to an external magnetic field); $\mathcal M = \gamma \mathcal J$ is magnetization length for one sublattice and $\gamma$ is the gyromagnetic ratio. The term $|\mathcal M \mathbf m|^2/2\chi$ represents an energy penalty for creating uniform magnetization; $\chi$ is paramagnetic susceptibility. Lastly, $\mathcal U(\mathbf n)$ is the potential energy density written in terms of staggered magnetization; it includes contribution of exchange interactions, anisotropy etc. 

The equation of motion for uniform magnetization $\mathbf m$ is simple as the Lagrangian (\ref{eq:L-n-m}) only contains terms linear and quadratic in $\mathbf m$. However, we have to respect the constraint $\mathbf m \cdot \mathbf n = 0$ (\ref{eq:constraints}). For this reason, only $\mathbf h_\perp = \mathbf n \times (\mathbf h \times \mathbf n)$, the component of $\mathbf h$ transverse to $\mathbf n$, enters the result: 
\begin{equation}
\mathbf m = \chi [\mathcal J \dot{\mathbf n}\times \mathbf n 
		+ \mathcal M \mathbf n \times (\mathbf h \times \mathbf n)]/\mathcal M^2.
\end{equation}
Eliminating the subdominant field $\mathbf m$ from the Lagrangian (\ref{eq:L-n-m}) yields a Lagrangian for staggered magnetization alone: 
\begin{equation}
\mathcal L(\mathbf n) = 
	\frac{\rho (\dot{\mathbf n} - \gamma \mathbf h \times \mathbf n)^2}{2}  
	- \mathcal U(\mathbf n).
\label{eq:L-n}
\end{equation} 

The term $\rho |\dot{\mathbf n}|^2/2$ in Eq.~(\ref{eq:L-n}) is the kinetic energy of staggered magnetization and $\rho = \chi/\gamma^2$ is the density of inertia \cite{PhysRevLett.50.1153, PhysRep.194.117}. This term endows antiferromagnetic solitons with a mass. Suppose a soliton is parametrized by a set of collective coordinates $\mathbf q = \{q_1, q_2, \ldots\}$ such as the position of a domain wall, the coordinates of a vortex core etc. The variation of $\mathbf n$ in time is mediated by the change of these collective coordinates: $\dot{\mathbf n} = \dot{q}_i \partial \mathbf n/\partial q_i$. The soliton's kinetic energy is then $M_{ij} \dot{q}_i \dot{q}_j/2$, where $M_{ij} = \rho \int dV \, \frac{\partial \mathbf n}{\partial q_i} \cdot \frac{\partial \mathbf n}{\partial q_j}$ is the inertia tensor \cite{PhysRevLett.110.127208}. 

The potential term $\rho |\gamma\mathbf h \times \mathbf n|^2/2$ in Eq.~(\ref{eq:L-n}) expresses local anisotropy favoring the direction of $\mathbf n$ orthogonal to the effective field $\mathbf h$. This term modifies the potential landscape $U(\mathbf q)$ of a soliton: 
\begin{equation}
U[\mathbf q, \mathbf h(\mathbf r)] = U[\mathbf q, 0] 
	-  \int dV \, \frac{\rho|\mathbf h \times \mathbf n|^2}{2}.
\label{eq:U-h}
\end{equation}

The cross term $\rho \gamma \mathbf h \cdot(\dot{\mathbf n} \times \mathbf n)$ in Eq.~(\ref{eq:L-n}) is linear in the time derivative $\dot{\mathbf n}$ and thus quantifies the effective geometric phase for the dynamics of staggered magnetization. In the Lagrangian of a soliton, it turns into $A_i \dot{q}_i$, a coupling to an external gauge field 
\begin{equation}
A_i(\mathbf q) = \int dV \, \rho \gamma \mathbf h \cdot 
	\left(
		\frac{\partial \mathbf n}{\partial q_i} \times \mathbf n
	\right).
\label{eq:A} 
\end{equation}

The equations of motion for an antiferromagnetic soliton have the form of Newton's second law for a particle of unit electric charge in this gauge field: 
\begin{equation}
M_{ij} \ddot{q}_j = - \partial U/\partial q_i + E_i + F_{ij} \dot{q}_i - M_{ij} \dot{q}_j/T.
\label{eq:2nd-law}
\end{equation}
The ``magnetic field'' $F_{ij} = -F_{ji}$ is the curl of the gauge potential: 
\begin{equation}
F_{ij} = \frac{\partial A_j}{\partial q_i} - \frac{\partial A_i}{\partial q_j}
	= - 2 \int dV \, \rho \gamma \mathbf h \cdot 
		\left(
			\frac{\partial \mathbf n}{\partial q_i} 
			\times 
			\frac{\partial \mathbf n}{\partial q_j}
		\right).
\label{eq:F-def}
\end{equation}
For the collective coordinates $X_\alpha$ representing rigid translations $x_\alpha \mapsto x_\alpha + X_\alpha$ of a magnetic soliton, the ``magnetic field'' $F_{\alpha\beta}$ is related to the gyrovector $\mathbf G$: $G_\alpha = \frac{1}{2}\epsilon_{\alpha\beta\gamma} F_{\beta\gamma}$; $F_\alpha = F_{\alpha\beta} \dot{X}_\alpha = \epsilon_{\alpha\beta\gamma} \dot{X}_\beta G_\gamma$ is the gyrotropic force \cite{PhysRevLett.30.230}. 
The ``electric field'' 
\begin{equation}
E_i = - \int dV \, \rho \gamma \dot{\mathbf h} \cdot 
	\left( \frac{\partial \mathbf n}{\partial q_i} \times \mathbf n \right)
\end{equation}
arises if $\mathbf h$ depends explicitly on time. 

The ``electromagnetic fields'' satisfy Jacobi identities
\begin{equation}
\frac{\partial E_j}{\partial q_i} 
- \frac{\partial E_i}{\partial q_j}
+ \frac{\partial F_{ij}}{\partial t} = 0,
\quad
\frac{\partial F_{ij}}{\partial q_k}
+ \frac{\partial F_{jk}}{\partial q_i}
+ \frac{\partial F_{ki}}{\partial q_j} = 0.
\end{equation}
the analogs of Maxwell's $\nabla \times \mathbf E + \dot{\mathbf B} = 0$ and $\nabla \cdot \mathbf B = 0$. In fact, we can define local versions of the ``electromagnetic fields'' as it was previously done for a ferromagnet \cite{JPCM.20.L83}, 
\begin{eqnarray}
A_\alpha &=& \rho \gamma \mathbf h
	\cdot (\partial_\alpha \mathbf n \times \mathbf n),
\nonumber\\
E_\alpha &=& - \rho \gamma \dot{\mathbf h} 
	\cdot (\partial_\alpha \mathbf n \times \mathbf n),
\\ 
B_\alpha &=& - \epsilon_{\alpha\beta\gamma} \, \rho \gamma \mathbf h
	\cdot (\partial_\beta \mathbf n \times \partial_\gamma \mathbf n).
\nonumber
\end{eqnarray}
The emergent fields couple to an electric current and are, in principle, measurable as in the ferromagnetic case \cite{PhysRevLett.102.067201}.

The last term on the right-hand side of Eq.~(\ref{eq:2nd-law}) is a viscous force with the mode-independent relaxation time $T = \rho/(2\alpha \mathcal J)$, where $\alpha$ is Gilbert's dimensionless damping constant \cite{PhysRevB.90.104406}. 

\emph{Domain wall in an easy-axis antiferromagnet.} We illustrate these general considerations on the familiar example of an easy-axis antiferromagnet in one dimension with potential energy density 
\begin{equation}
\mathcal U(\mathbf n) =  
		\frac{A}{2} \left|\frac{\partial \mathbf n}{\partial z}\right|^2
		+ \frac{K}{2} |\mathbf e_3 \times \mathbf n|^2.
\end{equation}
Here $A>0$ is the strength of exchange, $K>0$ is the anisotropy constant, and $\mathbf e_3 = (0,0,1)$. This system has two uniform ground states $\mathbf n = \pm \mathbf e_3$, linear excitations in the form of spin waves with the dispersion $\omega^2 = (K + Ak^2)/\rho$, and nonlinear solitons in the form of domain walls. Static domain walls have width $\lambda = \sqrt{A/K}$ and are parametrized in spherical angles $\theta(z)$ and $\phi(z)$ as follows: 
\begin{equation}
\cos{\theta(z)} = \pm \tanh{\frac{z-Z}{\lambda}}, 
\quad 
\phi(z) = \Phi. 
\label{eq:domain-wall}
\end{equation}
Position $Z$ and azimuthal angle $\Phi$ represent the two zero modes of the system associated with the global symmetries of translation and rotation. Weak external perturbations do not alter the shape of the soliton significantly and mostly induce the dynamics of $Z$ and $\Phi$. The Lagrangian of a domain wall at this level contains kinetic energy: $L = M \dot{Z}^2/2 + I \dot{\Phi}^2/2$, where $M = 2\rho/\lambda$ is the mass and $I = M \lambda^2$ is the moment of inertia. Thus a domain wall behaves like a point mass constrained to move on the surface of a cylinder of radius $\lambda$. 

In the simplest case, the linear in $\mathbf m$ term in Eq.~(\ref{eq:L-n-m}) comes from the external magnetic field $\mathbf H$, so that $\mathbf h = \mathbf H$. The gauge potential (\ref{eq:A}) for a domain wall (\ref{eq:domain-wall}) is 
\begin{equation}
A_Z = \pm \pi \rho \gamma (H_x \sin{\Phi} - H_y \cos{\Phi}), 
\quad
A_\Phi = - 2 \rho \lambda \gamma H_z.
\label{eq:A-Z-Phi}
\end{equation}
For a particle on the surface of a cylinder, these describe a ``magnetic field'' embedded in three dimensions,
\begin{equation}
\mathbf B = \frac{M \gamma}{2} (\pm \pi H_x, \pm \pi H_y, -4 H_z).
\end{equation}
When $\mathbf B$ is time-dependent, it induces an ``electric field'' $\mathbf E$ with the following axial and azimuthal components on the surface of the cylinder:
\begin{equation}
\mathbf E \cdot \mathbf e_3 = 
	\pm \frac{\pi M \lambda \gamma}{2} \dot{\mathbf H} \cdot \mathbf e_\phi,
\quad
\mathbf E \cdot \mathbf e_\phi = 
	M \lambda \gamma \dot{\mathbf H} \cdot \mathbf e_3,
\end{equation}
where $\mathbf e_\phi = (-\sin{\Phi}, \cos{\Phi}, 0)$ is a unit vector in the azimuthal direction.  The net ``electromagnetic'' force in the axial direction is 
\begin{equation}
F_Z^\mathrm{em} = E_Z + F_{Z\Phi}\dot{\Phi} 
	= \frac{d}{d t}
		\left(
			\pm \frac{\pi M \lambda \gamma}{2} \mathbf H \cdot \mathbf e_\phi
		\right).
\label{eq:Fem-Z}
\end{equation}

A sustained ``electromagnetic'' force can be generated if the real magnetic field $\mathbf H$ (more precisely, its azimuthal component $\mathbf H \cdot \mathbf e_\phi$) rises linearly in time. This is not a practical way to propel a domain wall. The ``electromagnetic'' force from an oscillating external field $\mathbf H(t)$ averages out to zero over time. To overcome this problem, \textcite{ApplPhysLett.109.122404} proposed a ratchet propulsion mechanism combining periodic field pulses with an asymmetric profile $\mathbf H(t)$ and static friction. If the field is ramped up and down at different rates, the friction force, opposing the motion of the domain wall, has different magnitudes during the rise and fall of the field pulse $\mathbf H(t)$. As a result, even though the average ``electromagnetic'' force vanishes, the friction force does not. 

The peculiar result for the ``electromagnetic'' force (\ref{eq:Fem-Z}) is not specific to the example of a domain wall. Generally, if a soliton has a zero mode $q_a$ associated with a global symmetry and the effective field $\mathbf h$ respects this symmetry, the corresponding ``electromagnetic'' force is given by the ``electric field'' alone:
\begin{equation}
F_a^\mathrm{em} 
	= \frac{\delta}{\delta q_a} \int dt\, A_i \dot{q}_i
	=  -\frac{dA_a}{dt} + \frac{\partial A_i}{\partial q_a}\dot{q}_i 
	=-\frac{d A_a}{d t}
\end{equation}
(translations in $q_a$ do not change gauge potentials $A_i$). The long-time average of the force is 0, unless $A_a(t)$ keeps growing in time. It would be interesting to explore whether a spatially nonuniform and time-dependent oscillating magnetic field $\mathbf H(\mathbf r, t)$ can be used to accelerate solitons.

Another important external perturbation is the spin torque from an electric current in a metallic antiferromagnet. Spins of electrons moving in an inhomogeneous magnetic background undergo precession and thus exchange angular momentum with the soliton. Here we focus on adiabatic spin torque that results when electron spins follow the local direction of magnetization. We rely on a simple model in which a conduction electron moves on one antiferromagnetic sublattice \cite{PhysRevLett.106.107206}. The effect of spin torque is computed for each sublattice independently and can be incorporated through a simple modification of the kinetic term in the Lagrangian: the time derivative $\partial_t$ is replaced with the convective derivative $\partial_t + \mathbf u \cdot \nabla$ \cite{PhysRep.468.213}. Here $\mathbf u$ is the drift velocity of electrons related to the electric current $\mathbf j = e n \mathbf u$; $n$ is the concentration of electrons. The Lagrangian density (\ref{eq:L-n-m}) acquires a term $\mathcal J \mathbf m \cdot [(\mathbf u \cdot \nabla) \mathbf n \times \mathbf n]$, from which we read off the effective magnetic field $\mathbf h = \gamma^{-1} (\mathbf u \cdot \nabla) \mathbf n \times \mathbf n$. The induced uniform magnetization $\mathcal M \mathbf m = \gamma \rho (\mathbf u \cdot \nabla) \mathbf n \times \mathbf n$ agrees with the standard phenomenology of adiabatic spin torque \cite{PhysRevLett.106.107206}.

Returning to our model of an easy-axis antiferromagnet in one dimension, we compute the gauge potential (\ref{eq:A}) with $\gamma \mathbf h = u \, \partial_z \mathbf n \times \mathbf n$ to obtain 
\begin{equation}
A_Z = - M u, 
\quad 
A_\Phi = 0.
\end{equation}
The ``magnetic field'' $F_{Z\Phi} = \partial_Z A_\Phi - \partial_\Phi A_Z = 0$, whereas the ``electric field'' $E_Z = - \dot{A}_Z = M\dot{u}$ is once again proportional to the time derivative of an external perturbation. Thus adiabatic spin torque cannot be used to propel a domain wall \cite{PhysRevLett.106.107206}. 

\emph{Vortex in an easy-plane antiferromagnet.} Consider a Heisenberg antiferromagnet in two spatial dimensions with easy-plane ($K<0$) anisotropy with potential energy density
\begin{equation}
\mathcal U(\mathbf n) =  
		\frac{A}{2} \left|\nabla \mathbf n\right|^2
		+ \frac{K}{2} |\mathbf e_3 \times \mathbf n|^2.
\end{equation}
It has uniform ground states $\mathbf n = (\cos{\phi}, \sin{\phi}, 0)$. Topological solitons are vortices $\mathbf n(\mathbf r - \mathbf R)$, where $\mathbf R = (X, Y)$ is the center of the vortex. A vortex centered at the origin, $\mathbf n(\mathbf r)$, is parametrized in spherical angles as
\begin{equation}
e^{i\phi(\mathbf r)}  
	= \left(\frac{x + i y}{|x + i y|}\right)^n,
\quad
\cos{\theta(\mathbf r)} = \pm f_n(r/\lambda).
\label{eq:phi-vortex}
\end{equation}
Here $n \in \mathbb Z$ is the vortex winding number. The function $f(\xi)$ is a profile of the out-of-plane magnetization at the vortex core with $f_n(0) = 1$ and $f_n(\infty) = 0$; $\lambda = \sqrt{A/|K|}$ is the radius of the core. The vortex mass $M = \pi \rho \ln{(\Lambda/\lambda)}$ depends logarithmically on the core radius $\lambda$ and on a long-distance cutoff $\Lambda$, which can be the size of the system or the screening length due to the presence of other vortices. 

A magnetic field $\mathbf H = H \mathbf e_3$ along the hard axis breaks the time-reversal symmetry of the antiferromagnet and allows for non-vanishing gyrotropic coefficients 
\begin{equation}
F_{XY} = - F_{YX} = \int d^2r \, (-2\rho \gamma \mathbf H) \cdot 
		(\partial_x \mathbf n \times \partial_y \mathbf n).
\label{eq:F-XY-1}
\end{equation}
This expression follows from Eq.~(\ref{eq:F-def}) under the assumption of a rigid soliton $\mathbf n(\mathbf r - \mathbf R)$, for which $\partial_X = - \partial_x$ and $\partial_Y = - \partial_y$. To bring out the topological nature of this quantity, we recast the integrand as a curl $\partial_x a_y - \partial_y a_x$ of the vector $a_\alpha = \rho \gamma \mathbf H \cdot (\partial_\alpha \mathbf n \times \mathbf n)$ and use Stokes' theorem to transform the area integral (\ref{eq:F-XY-1}) into a line integral $\oint dx_\alpha \, \rho \gamma \mathbf H \cdot (\partial_\alpha \mathbf n \times \mathbf n)$ over the boundary. Away from the vortex core, $\mathbf n$ is in the easy plane, $\theta=\pi/2$, and $\mathbf H \cdot (\partial_\alpha \mathbf n \times \mathbf n) = - H \partial_\alpha \phi$. Hence the gyrotropic coefficients of a vortex,
\begin{equation}
F_{XY} = -F_{YX} 
	= - \rho \gamma H \oint dx_\alpha \, \partial_\alpha \phi
	= - 2\pi n \rho \gamma H. 
\label{eq:F-XY-H}
\end{equation}
This result was first obtained by \textcite{PhysRevLett.72.404}. 

\begin{figure}
\includegraphics[width=0.98\columnwidth]{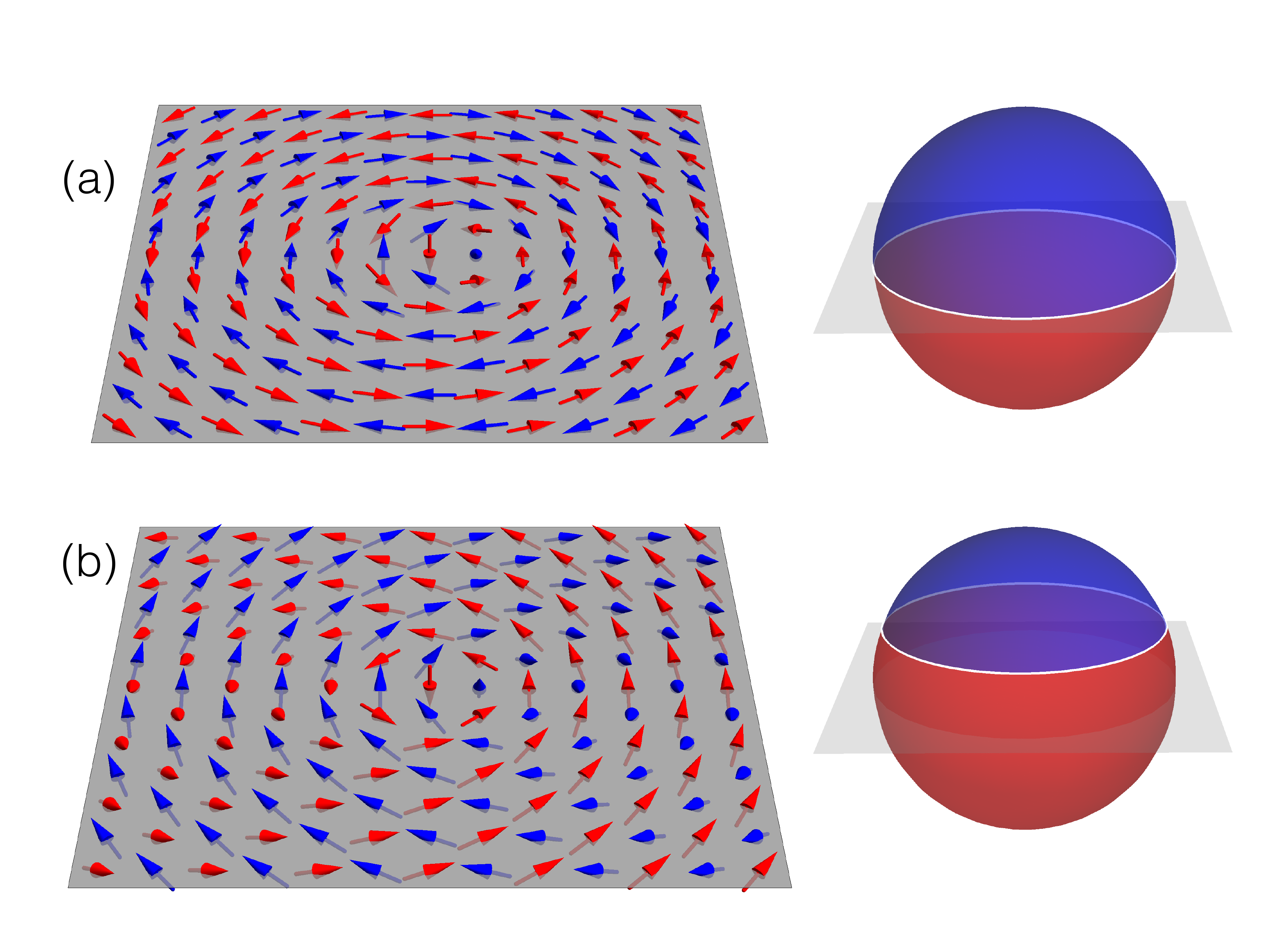}
\caption{A vortex in an antiferromagnet without an applied field (a) and in a weak field $\mathbf H = (0,0,H)$. Sublattice magnetizations $\mathbf m_1$ and $\mathbf m_2$ are shown in blue and red. The right panels show portions of the unit sphere covered by the magnetization fields $\mathbf m_1(\mathbf r)$ and $\mathbf m_2(\mathbf r)$. }
\label{fig:vortex}
\end{figure}

The topological nature of the gyrotropic coefficients (\ref{eq:F-XY-H}) clearly comes into focus if we view a vortex in the two antiferromagnetic sublattices separately, as if they were two independent ferromagnets. In the absence of an applied field, sublattice magnetizations $\mathbf m_1(\mathbf r)$ and $\mathbf m_2(\mathbf r)$ point in opposite directions and cover the northern and southern hemispheres, Fig.~\ref{fig:vortex}(a). This endows them with equal and opposite skyrmion numbers $q = \pm n/2$ and gyrotropic coefficients $F_{XY} = 4\pi q \mathcal J = \pm 2\pi n \mathcal J$ \cite{PhysRep.468.213, PhysRevLett.100.127204}. The net gyrotropic coefficient is zero. In an applied magnetic field, both magnetizations tilt out of the easy plane toward the north pole by a small angle $\delta \theta = \frac{\chi H}{2\mathcal M} = \frac{\rho \gamma H}{2\mathcal J}$. Now $\mathbf m_1$ covers slightly less than the northern hemisphere and $\mathbf m_2$ slightly more than the southern hemisphere, Fig.~\ref{fig:vortex}(b). The respective skyrmion charges are $q = \pm \frac{n}{2} - \frac{n \rho \gamma H}{4 \mathcal J}$. The net gyrotropic coefficient is then $F_{XY} = - 2\pi n \rho \gamma H$. 

In a two-dimensional ferromagnet, the gyrotropic tensor $F_{\alpha\beta}$ quantifies not only the Lorentz force $F_\alpha = F_{\alpha\beta} \dot{X}_\beta$ acting on a moving vortex, but also the Magnus force $F_\alpha = - F_{\alpha\beta} u_\beta$ exerted on the vortex core by a spin-polarized current of electrons flowing at the drift velocity $\mathbf u$ \cite{PhysRep.468.213}. It is reasonable to expect the same from our antiferromagnet. 

The interplay of adiabatic spin torque and an external magnetic field occurs through the potential term in the Lagrangian density $ \rho|\gamma \mathbf h \times \mathbf n|^2/2$ with the effective field $\mathcal \gamma \mathbf h = \gamma \mathbf H + (\mathbf u \cdot \nabla) \mathbf n \times \mathbf n$, namely through the part $\rho \gamma \mathbf H \cdot (u_\alpha \partial_\alpha \mathbf n \times \mathbf n)$ that is linear in the applied field $\mathbf H$ and the drift velocity $\mathbf u$. Its contribution to the stress tensor is
\begin{equation}
\sigma_{\alpha\beta} 
	= \epsilon_{\alpha\mu} \epsilon_{\beta\nu} u_\mu \rho \gamma \mathbf H 
		\cdot (\partial_\nu \mathbf n \times \mathbf n)
\end{equation}
in 2 spatial dimensions. The Magnus force on the vortex core is obtained by integrating stress around a contour containing the core, $F_\alpha = - \oint \sigma_{\alpha \beta} \, dS_\beta$, where $dS_\beta = \epsilon_{\beta \lambda} dx_\lambda$ is an ``area'' element normal to the contour segment $dx_\lambda$: 
\begin{eqnarray}
F_\alpha &=& 
- \epsilon_{\alpha\mu} u_\mu 
		\oint dx_\nu \, \rho \gamma \mathbf H 
			\cdot (\partial_\nu \mathbf n \times \mathbf n)
\nonumber\\
&=& \epsilon_{\alpha \mu} u_\mu \rho \gamma H \oint dx_\nu \, \partial_\nu \phi
	= - F_{\alpha \mu} u_\mu,
\end{eqnarray}
as expected. In 3 dimensions, this translates into Eq.~(\ref{eq:Magnus-force}). 

In a weak magnetic field, the velocity of the vortex $\mathbf v$ is set by the balance between the Magnus force and the viscous force $- M \mathbf v/T$, so that $\mathbf v$ is orthogonal to $\mathbf u$ and their magnitudes are related by $v \approx \gamma H T u$. In a strong field, the gyrotropic force becomes dominant and $\mathbf v$ approaches $\mathbf u$. The crossover field being $H_\mathrm{cr} \approx 1/(\gamma T)$. In an insulating antiferromagnet Cr$_2$O$_3$, $\gamma = 1.76 \times 10^{11}$ s$^{-1}$ T$^{-1}$ and $T = 60$ ps \cite{ApplPhysLett.108.132403}, so $H_\mathrm{cr} \approx 0.1$ T. In metallic antiferromagnets, the relaxation time $T$ is expected to be shorter and the crossover field higher. The spin drift velocity $u$ is of the order of 5 m/s for a current density $j = 10^{11}$ A/m$^2$ \cite{PhysRevB.81.140407}. 

We thank Yaroslav Tserkovnyak and Jiadong Zang for useful discussions. Research was supported by the U.S. Department of Energy, Office of Basic Energy Sciences, Division of Materials Sciences and Engineering under Award DE-FG02-08ER46544 and by the U.S. Army Research Office under Contract No. 911NF-14-1-0016.

\bibliographystyle{apsrev4-1}
\bibliography{afm-solitons}

\begin{thebibliography}{21}%
\makeatletter
\providecommand \@ifxundefined [1]{%
 \@ifx{#1\undefined}
}%
\providecommand \@ifnum [1]{%
 \ifnum #1\expandafter \@firstoftwo
 \else \expandafter \@secondoftwo
 \fi
}%
\providecommand \@ifx [1]{%
 \ifx #1\expandafter \@firstoftwo
 \else \expandafter \@secondoftwo
 \fi
}%
\providecommand \natexlab [1]{#1}%
\providecommand \enquote  [1]{``#1''}%
\providecommand \bibnamefont  [1]{#1}%
\providecommand \bibfnamefont [1]{#1}%
\providecommand \citenamefont [1]{#1}%
\providecommand \href@noop [0]{\@secondoftwo}%
\providecommand \href [0]{\begingroup \@sanitize@url \@href}%
\providecommand \@href[1]{\@@startlink{#1}\@@href}%
\providecommand \@@href[1]{\endgroup#1\@@endlink}%
\providecommand \@sanitize@url [0]{\catcode `\\12\catcode `\$12\catcode
  `\&12\catcode `\#12\catcode `\^12\catcode `\_12\catcode `\%12\relax}%
\providecommand \@@startlink[1]{}%
\providecommand \@@endlink[0]{}%
\providecommand \url  [0]{\begingroup\@sanitize@url \@url }%
\providecommand \@url [1]{\endgroup\@href {#1}{\urlprefix }}%
\providecommand \urlprefix  [0]{URL }%
\providecommand \Eprint [0]{\href }%
\providecommand \doibase [0]{http://dx.doi.org/}%
\providecommand \selectlanguage [0]{\@gobble}%
\providecommand \bibinfo  [0]{\@secondoftwo}%
\providecommand \bibfield  [0]{\@secondoftwo}%
\providecommand \translation [1]{[#1]}%
\providecommand \BibitemOpen [0]{}%
\providecommand \bibitemStop [0]{}%
\providecommand \bibitemNoStop [0]{.\EOS\space}%
\providecommand \EOS [0]{\spacefactor3000\relax}%
\providecommand \BibitemShut  [1]{\csname bibitem#1\endcsname}%
\let\auto@bib@innerbib\@empty
\bibitem [{\citenamefont {Kosevich}\ \emph {et~al.}(1990)\citenamefont
  {Kosevich}, \citenamefont {Ivanov},\ and\ \citenamefont
  {Kovalev}}]{PhysRep.194.117}%
  \BibitemOpen
  \bibfield  {author} {\bibinfo {author} {\bibfnamefont {A.~M.}\ \bibnamefont
  {Kosevich}}, \bibinfo {author} {\bibfnamefont {B.~A.}\ \bibnamefont
  {Ivanov}}, \ and\ \bibinfo {author} {\bibfnamefont {A.~S.}\ \bibnamefont
  {Kovalev}},\ }\href {\doibase 10.1016/0370-1573(90)90130-T} {\bibfield
  {journal} {\bibinfo  {journal} {Phys. Rep.}\ }\textbf {\bibinfo {volume}
  {194}},\ \bibinfo {pages} {117} (\bibinfo {year} {1990})}\BibitemShut
  {NoStop}%
\bibitem [{\citenamefont {Parkin}\ \emph {et~al.}(2008)\citenamefont {Parkin},
  \citenamefont {Hayashi},\ and\ \citenamefont {Thomas}}]{Science.320.5873}%
  \BibitemOpen
  \bibfield  {author} {\bibinfo {author} {\bibfnamefont {S.~S.~P.}\
  \bibnamefont {Parkin}}, \bibinfo {author} {\bibfnamefont {M.}~\bibnamefont
  {Hayashi}}, \ and\ \bibinfo {author} {\bibfnamefont {L.}~\bibnamefont
  {Thomas}},\ }\href {\doibase 10.1126/science.1145799} {\bibfield  {journal}
  {\bibinfo  {journal} {Science}\ }\textbf {\bibinfo {volume} {320}},\ \bibinfo
  {pages} {190} (\bibinfo {year} {2008})}\BibitemShut {NoStop}%
\bibitem [{\citenamefont {Nagaosa}\ and\ \citenamefont
  {Tokura}(2013)}]{NatNano.8.899}%
  \BibitemOpen
  \bibfield  {author} {\bibinfo {author} {\bibfnamefont {N.}~\bibnamefont
  {Nagaosa}}\ and\ \bibinfo {author} {\bibfnamefont {Y.}~\bibnamefont
  {Tokura}},\ }\href {\doibase 10.1038/nnano.2013.243} {\bibfield  {journal}
  {\bibinfo  {journal} {Nat. Nano.}\ }\textbf {\bibinfo {volume} {8}},\
  \bibinfo {pages} {899} (\bibinfo {year} {2013})}\BibitemShut {NoStop}%
\bibitem [{\citenamefont {Jungwirth}\ \emph {et~al.}(2016)\citenamefont
  {Jungwirth}, \citenamefont {Marti}, \citenamefont {Wadley},\ and\
  \citenamefont {Wunderlich}}]{NatNano.11.231}%
  \BibitemOpen
  \bibfield  {author} {\bibinfo {author} {\bibfnamefont {T.}~\bibnamefont
  {Jungwirth}}, \bibinfo {author} {\bibfnamefont {X.}~\bibnamefont {Marti}},
  \bibinfo {author} {\bibfnamefont {P.}~\bibnamefont {Wadley}}, \ and\ \bibinfo
  {author} {\bibfnamefont {J.}~\bibnamefont {Wunderlich}},\ }\href {\doibase
  10.1038/nnano.2016.18} {\bibfield  {journal} {\bibinfo  {journal} {Nat.
  Nano.}\ }\textbf {\bibinfo {volume} {11}},\ \bibinfo {pages} {231} (\bibinfo
  {year} {2016})}\BibitemShut {NoStop}%
\bibitem [{\citenamefont {Dzyaloshinsky}(1958)}]{JPhysChemSol.4.241}%
  \BibitemOpen
  \bibfield  {author} {\bibinfo {author} {\bibfnamefont {I.}~\bibnamefont
  {Dzyaloshinsky}},\ }\href {\doibase 10.1016/0022-3697(58)90076-3} {\bibfield
  {journal} {\bibinfo  {journal} {J. Phys. Chem. Sol.}\ }\textbf {\bibinfo
  {volume} {4}},\ \bibinfo {pages} {241} (\bibinfo {year} {1958})}\BibitemShut
  {NoStop}%
\bibitem [{\citenamefont {Belashchenko}\ \emph {et~al.}(2016)\citenamefont
  {Belashchenko}, \citenamefont {Tchernyshyov}, \citenamefont {Kovalev},\ and\
  \citenamefont {Tretiakov}}]{ApplPhysLett.108.132403}%
  \BibitemOpen
  \bibfield  {author} {\bibinfo {author} {\bibfnamefont {K.~D.}\ \bibnamefont
  {Belashchenko}}, \bibinfo {author} {\bibfnamefont {O.}~\bibnamefont
  {Tchernyshyov}}, \bibinfo {author} {\bibfnamefont {A.~A.}\ \bibnamefont
  {Kovalev}}, \ and\ \bibinfo {author} {\bibfnamefont {O.~A.}\ \bibnamefont
  {Tretiakov}},\ }\href {\doibase 10.1063/1.4944996} {\bibfield  {journal}
  {\bibinfo  {journal} {Appl. Phys. Lett.}\ }\textbf {\bibinfo {volume}
  {108}},\ \bibinfo {pages} {132403} (\bibinfo {year} {2016})}\BibitemShut
  {NoStop}%
\bibitem [{\citenamefont {Hals}\ \emph {et~al.}(2011)\citenamefont {Hals},
  \citenamefont {Tserkovnyak},\ and\ \citenamefont
  {Brataas}}]{PhysRevLett.106.107206}%
  \BibitemOpen
  \bibfield  {author} {\bibinfo {author} {\bibfnamefont {K.~M.~D.}\
  \bibnamefont {Hals}}, \bibinfo {author} {\bibfnamefont {Y.}~\bibnamefont
  {Tserkovnyak}}, \ and\ \bibinfo {author} {\bibfnamefont {A.}~\bibnamefont
  {Brataas}},\ }\href {\doibase 10.1103/PhysRevLett.106.107206} {\bibfield
  {journal} {\bibinfo  {journal} {Phys. Rev. Lett.}\ }\textbf {\bibinfo
  {volume} {106}},\ \bibinfo {pages} {107206} (\bibinfo {year}
  {2011})}\BibitemShut {NoStop}%
\bibitem [{\citenamefont {Gomonay}\ \emph {et~al.}(2016)\citenamefont
  {Gomonay}, \citenamefont {Kl{\"a}ui},\ and\ \citenamefont
  {Sinova}}]{ApplPhysLett.109.122404}%
  \BibitemOpen
  \bibfield  {author} {\bibinfo {author} {\bibfnamefont {O.}~\bibnamefont
  {Gomonay}}, \bibinfo {author} {\bibfnamefont {M.}~\bibnamefont {Kl{\"a}ui}},
  \ and\ \bibinfo {author} {\bibfnamefont {J.}~\bibnamefont {Sinova}},\ }\href
  {\doibase 10.1063/1.4964272} {\bibfield  {journal} {\bibinfo  {journal}
  {Appl. Phys. Lett.}\ }\textbf {\bibinfo {volume} {109}},\ \bibinfo {pages}
  {142404} (\bibinfo {year} {2016})}\BibitemShut {NoStop}%
\bibitem [{\citenamefont {De~Gennes}\ and\ \citenamefont
  {Matricon}(1964)}]{Rev.Mod.Phys.36.45}%
  \BibitemOpen
  \bibfield  {author} {\bibinfo {author} {\bibfnamefont {P.~G.}\ \bibnamefont
  {De~Gennes}}\ and\ \bibinfo {author} {\bibfnamefont {J.}~\bibnamefont
  {Matricon}},\ }\href {\doibase 10.1103/RevModPhys.36.45} {\bibfield
  {journal} {\bibinfo  {journal} {Rev. Mod. Phys.}\ }\textbf {\bibinfo {volume}
  {36}},\ \bibinfo {pages} {45} (\bibinfo {year} {1964})}\BibitemShut {NoStop}%
\bibitem [{\citenamefont {Ivanov}\ and\ \citenamefont
  {Kolezhuk}(1995)}]{PhysRevLett.74.1859}%
  \BibitemOpen
  \bibfield  {author} {\bibinfo {author} {\bibfnamefont {B.~A.}\ \bibnamefont
  {Ivanov}}\ and\ \bibinfo {author} {\bibfnamefont {A.~K.}\ \bibnamefont
  {Kolezhuk}},\ }\href {\doibase 10.1103/PhysRevLett.74.1859} {\bibfield
  {journal} {\bibinfo  {journal} {Phys. Rev. Lett.}\ }\textbf {\bibinfo
  {volume} {74}},\ \bibinfo {pages} {1859} (\bibinfo {year}
  {1995})}\BibitemShut {NoStop}%
\bibitem [{\citenamefont {Kim}\ \emph {et~al.}(2014)\citenamefont {Kim},
  \citenamefont {Tserkovnyak},\ and\ \citenamefont
  {Tchernyshyov}}]{PhysRevB.90.104406}%
  \BibitemOpen
  \bibfield  {author} {\bibinfo {author} {\bibfnamefont {S.~K.}\ \bibnamefont
  {Kim}}, \bibinfo {author} {\bibfnamefont {Y.}~\bibnamefont {Tserkovnyak}}, \
  and\ \bibinfo {author} {\bibfnamefont {O.}~\bibnamefont {Tchernyshyov}},\
  }\href {\doibase 10.1103/PhysRevB.90.104406} {\bibfield  {journal} {\bibinfo
  {journal} {Phys. Rev. B}\ }\textbf {\bibinfo {volume} {90}},\ \bibinfo
  {pages} {104406} (\bibinfo {year} {2014})}\BibitemShut {NoStop}%
\bibitem [{\citenamefont {Tveten}\ \emph {et~al.}(2016)\citenamefont {Tveten},
  \citenamefont {M\"uller}, \citenamefont {Linder},\ and\ \citenamefont
  {Brataas}}]{PhysRevB.93.104408}%
  \BibitemOpen
  \bibfield  {author} {\bibinfo {author} {\bibfnamefont {E.~G.}\ \bibnamefont
  {Tveten}}, \bibinfo {author} {\bibfnamefont {T.}~\bibnamefont {M\"uller}},
  \bibinfo {author} {\bibfnamefont {J.}~\bibnamefont {Linder}}, \ and\ \bibinfo
  {author} {\bibfnamefont {A.}~\bibnamefont {Brataas}},\ }\href {\doibase
  10.1103/PhysRevB.93.104408} {\bibfield  {journal} {\bibinfo  {journal} {Phys.
  Rev. B}\ }\textbf {\bibinfo {volume} {93}},\ \bibinfo {pages} {104408}
  (\bibinfo {year} {2016})}\BibitemShut {NoStop}%
\bibitem [{\citenamefont {Haldane}(1983)}]{PhysRevLett.50.1153}%
  \BibitemOpen
  \bibfield  {author} {\bibinfo {author} {\bibfnamefont {F.~D.~M.}\
  \bibnamefont {Haldane}},\ }\href {\doibase 10.1103/PhysRevLett.50.1153}
  {\bibfield  {journal} {\bibinfo  {journal} {Phys. Rev. Lett.}\ }\textbf
  {\bibinfo {volume} {50}},\ \bibinfo {pages} {1153} (\bibinfo {year}
  {1983})}\BibitemShut {NoStop}%
\bibitem [{\citenamefont {Tveten}\ \emph {et~al.}(2013)\citenamefont {Tveten},
  \citenamefont {Qaiumzadeh}, \citenamefont {Tretiakov},\ and\ \citenamefont
  {Brataas}}]{PhysRevLett.110.127208}%
  \BibitemOpen
  \bibfield  {author} {\bibinfo {author} {\bibfnamefont {E.~G.}\ \bibnamefont
  {Tveten}}, \bibinfo {author} {\bibfnamefont {A.}~\bibnamefont {Qaiumzadeh}},
  \bibinfo {author} {\bibfnamefont {O.~A.}\ \bibnamefont {Tretiakov}}, \ and\
  \bibinfo {author} {\bibfnamefont {A.}~\bibnamefont {Brataas}},\ }\href
  {\doibase 10.1103/PhysRevLett.110.127208} {\bibfield  {journal} {\bibinfo
  {journal} {Phys. Rev. Lett.}\ }\textbf {\bibinfo {volume} {110}},\ \bibinfo
  {pages} {127208} (\bibinfo {year} {2013})}\BibitemShut {NoStop}%
\bibitem [{\citenamefont {Thiele}(1973)}]{PhysRevLett.30.230}%
  \BibitemOpen
  \bibfield  {author} {\bibinfo {author} {\bibfnamefont {A.~A.}\ \bibnamefont
  {Thiele}},\ }\href {\doibase 10.1103/PhysRevLett.30.230} {\bibfield
  {journal} {\bibinfo  {journal} {Phys. Rev. Lett.}\ }\textbf {\bibinfo
  {volume} {30}},\ \bibinfo {pages} {230} (\bibinfo {year} {1973})}\BibitemShut
  {NoStop}%
\bibitem [{\citenamefont {Volovik}(1987)}]{JPCM.20.L83}%
  \BibitemOpen
  \bibfield  {author} {\bibinfo {author} {\bibfnamefont {G.~E.}\ \bibnamefont
  {Volovik}},\ }\href {\doibase 10.1088/0022-3719/20/7/003} {\bibfield
  {journal} {\bibinfo  {journal} {J. Phys.: Condens. Matter}\ }\textbf
  {\bibinfo {volume} {20}},\ \bibinfo {pages} {L83} (\bibinfo {year}
  {1987})}\BibitemShut {NoStop}%
\bibitem [{\citenamefont {Yang}\ \emph {et~al.}(2009)\citenamefont {Yang},
  \citenamefont {Beach}, \citenamefont {Knutson}, \citenamefont {Xiao},
  \citenamefont {Niu}, \citenamefont {Tsoi},\ and\ \citenamefont
  {Erskine}}]{PhysRevLett.102.067201}%
  \BibitemOpen
  \bibfield  {author} {\bibinfo {author} {\bibfnamefont {S.~A.}\ \bibnamefont
  {Yang}}, \bibinfo {author} {\bibfnamefont {G.~S.~D.}\ \bibnamefont {Beach}},
  \bibinfo {author} {\bibfnamefont {C.}~\bibnamefont {Knutson}}, \bibinfo
  {author} {\bibfnamefont {D.}~\bibnamefont {Xiao}}, \bibinfo {author}
  {\bibfnamefont {Q.}~\bibnamefont {Niu}}, \bibinfo {author} {\bibfnamefont
  {M.}~\bibnamefont {Tsoi}}, \ and\ \bibinfo {author} {\bibfnamefont {J.~L.}\
  \bibnamefont {Erskine}},\ }\href {\doibase 10.1103/PhysRevLett.102.067201}
  {\bibfield  {journal} {\bibinfo  {journal} {Phys. Rev. Lett.}\ }\textbf
  {\bibinfo {volume} {102}},\ \bibinfo {pages} {067201} (\bibinfo {year}
  {2009})}\BibitemShut {NoStop}%
\bibitem [{\citenamefont {Tatara}\ \emph {et~al.}(2008)\citenamefont {Tatara},
  \citenamefont {Kohno},\ and\ \citenamefont {Shibata}}]{PhysRep.468.213}%
  \BibitemOpen
  \bibfield  {author} {\bibinfo {author} {\bibfnamefont {G.}~\bibnamefont
  {Tatara}}, \bibinfo {author} {\bibfnamefont {H.}~\bibnamefont {Kohno}}, \
  and\ \bibinfo {author} {\bibfnamefont {J.}~\bibnamefont {Shibata}},\ }\href
  {\doibase 10.1016/j.physrep.2008.07.003} {\bibfield  {journal} {\bibinfo
  {journal} {Phys. Rep.}\ }\textbf {\bibinfo {volume} {468}},\ \bibinfo {pages}
  {213} (\bibinfo {year} {2008})}\BibitemShut {NoStop}%
\bibitem [{\citenamefont {Ivanov}\ and\ \citenamefont
  {Sheka}(1994)}]{PhysRevLett.72.404}%
  \BibitemOpen
  \bibfield  {author} {\bibinfo {author} {\bibfnamefont {B.~A.}\ \bibnamefont
  {Ivanov}}\ and\ \bibinfo {author} {\bibfnamefont {D.~D.}\ \bibnamefont
  {Sheka}},\ }\href {\doibase 10.1103/PhysRevLett.72.404} {\bibfield  {journal}
  {\bibinfo  {journal} {Phys. Rev. Lett.}\ }\textbf {\bibinfo {volume} {72}},\
  \bibinfo {pages} {404} (\bibinfo {year} {1994})}\BibitemShut {NoStop}%
\bibitem [{\citenamefont {Tretiakov}\ \emph {et~al.}(2008)\citenamefont
  {Tretiakov}, \citenamefont {Clarke}, \citenamefont {Chern}, \citenamefont
  {Bazaliy},\ and\ \citenamefont {Tchernyshyov}}]{PhysRevLett.100.127204}%
  \BibitemOpen
  \bibfield  {author} {\bibinfo {author} {\bibfnamefont {O.~A.}\ \bibnamefont
  {Tretiakov}}, \bibinfo {author} {\bibfnamefont {D.}~\bibnamefont {Clarke}},
  \bibinfo {author} {\bibfnamefont {G.-W.}\ \bibnamefont {Chern}}, \bibinfo
  {author} {\bibfnamefont {Y.~B.}\ \bibnamefont {Bazaliy}}, \ and\ \bibinfo
  {author} {\bibfnamefont {O.}~\bibnamefont {Tchernyshyov}},\ }\href {\doibase
  10.1103/PhysRevLett.100.127204} {\bibfield  {journal} {\bibinfo  {journal}
  {Phys. Rev. Lett.}\ }\textbf {\bibinfo {volume} {100}},\ \bibinfo {pages}
  {127204} (\bibinfo {year} {2008})}\BibitemShut {NoStop}%
\bibitem [{\citenamefont {Zhu}\ \emph {et~al.}(2010)\citenamefont {Zhu},
  \citenamefont {Dennis},\ and\ \citenamefont
  {McMichael}}]{PhysRevB.81.140407}%
  \BibitemOpen
  \bibfield  {author} {\bibinfo {author} {\bibfnamefont {M.}~\bibnamefont
  {Zhu}}, \bibinfo {author} {\bibfnamefont {C.~L.}\ \bibnamefont {Dennis}}, \
  and\ \bibinfo {author} {\bibfnamefont {R.~D.}\ \bibnamefont {McMichael}},\
  }\href {\doibase 10.1103/PhysRevB.81.140407} {\bibfield  {journal} {\bibinfo
  {journal} {Phys. Rev. B}\ }\textbf {\bibinfo {volume} {81}},\ \bibinfo
  {pages} {140407} (\bibinfo {year} {2010})}\BibitemShut {NoStop}%
\end{thebibliography}%

\end{document}